\documentclass[epj]{svjour}
\usepackage{graphics}

\begin{document}
\title{Calculations of $^{6}$He+p elastic scattering cross sections using
folding approach and high-energy approximation for the optical potential}
\author{K.V. Lukyanov\inst{1}, V.K. Lukyanov\inst{1}, E.V. Zemlyanaya\inst{1},
A.N. Antonov\inst{2}, \and M.K. Gaidarov\inst{2,3} 
\thanks{\emph{e-mail: gaidarov@inrne.bas.bg}
Present address: Insert the address here if needed}%
}                     
%
%
\institute{Joint Institute for Nuclear Research, Dubna 141980,
Russia \and Institute for Nuclear Research and Nuclear Energy,
Bulgarian Academy of Sciences, Sofia 1784, Bulgaria \and Instituto
de Estructura de la Materia, CSIC, Serrano 123, E-28006 Madrid,
Spain}
\date{Received: date / Revised version: date}
%
\abstract{Calculations of microscopic optical potentials (OP's)
(their real and imaginary parts) are performed to analyze the
$^6$He+p elastic scattering data at a few tens of MeV/nucleon
(MeV/N). The OP's and the cross sections are calculated using
three model densities of $^6$He. Effects of the regularization of
the NN forces and their dependence on nuclear density are
investigated. Also, the role of the spin-orbit terms and of the
non-linearity in the calculations of the OP's, as well as effects
of their renormalization are studied. The sensitivity of the cross
sections to the nuclear densities was tested and one of them that
gives a better agreement with the data was chosen.
} 
\titlerunning{Calculations of $^{6}$He+p elastic scattering cross sections}
\authorrunning{K.V. Lukyanov {\it et al.}}

\maketitle
\section{Introduction}
\label{intro}

The basic characteristics of the exotic nuclei, such as their
charge and matter distributions  have been tested, in particular,
by studying differential and total reaction cross sections of the
proton scattering on exotic nuclei in inverse kinematics.
Nowadays, a substantial amount of experimental data exists for the
cross sections of $^6$He+p elastic scattering at different
energies. For example, proton elastic scattering angular
distributions were measured at incident energies less than 100
MeV/N for $^6$He, namely 25.2 \cite{[1],[2],[3],[4]}, 38.3
\cite{[5]}, 41.6 \cite{[6],[7],[8]}, and 71 MeV/N \cite{[9],[10]}
and at energy of 700 MeV/N for He and Li isotopes ($e.g.$ refs.
\cite{[11],[12],[13],[14],[15]}). The analyses of differential and
total reaction cross sections have been performed using different
phenomenological and microscopic methods and models of nuclear
structure (see, $e.g.$ refs.
\cite{[9],[10],[11],[12],[13],[14],[15],[16],[17],[18],[19],[20],[21]}).
We note also the microscopic analysis
(refs.~\cite{[22],[23],Amos2006,Deb2005,Deb2001,Deb2003}) based on
the "coordinate-space $g$-matrix folding method" \cite{Amos00}
(and some modifications \cite{Arell2007}), where non-local OP is
obtained using a folding of a local medium-dependent NN effective
interaction with the target ground-state mixed density. In some of
the calculations ($e.g.$ \cite{[9],[10],[16],[17]}) the eikonal
approach using proton and neutron density distributions, as well
as parametrized NN total cross section have been used. It has been
accepted that for energies larger than 500 MeV/N the multiple
scattering diffraction theory developed in \cite{{[47]},{[48]}}
(the Glauber theory) is a relevant method to study charge and
matter distributions from proton elastic scattering data
\cite{[11],[24],[25]}.

The experimental information on cross sections of\\ $^6$He+p
elastic scattering requires for its adequate description a
development of the respective microscopic methods  of their
analysis. These methods give an opportunity, first, to distinguish
between different models for the exotic $^6$He as a nucleus with a
halo with two neutrons and, secondly, to test the attempts of the
folding approach for constructions of optical potentials. The
latter include the understanding of the role of various components
of the OP and the necessity to introduce fitting parameters. A
number of works has been devoted to calculations of OP's using the
folding approach (see, $e.g.$
\cite{[20],[21],[22],[23]},\cite{[26],[27],[28],[29],[30]}). For
instance, the real parts of OP's for calculating the $^6$He+p,
$^6$He+ $^4$He ($E_{lab}$=151 MeV) \cite{[20]} and $^6$He+p,
$^8$He+p ($E_{inc}<100$ MeV) \cite{[21]} elastic differential
cross sections have been obtained microscopically using realistic
M3Y-Paris effective NN interaction \cite{[27],[30],[31]} together
with the Tanihata {\it et al.} proton and neutron densities of the
helium isotopes \cite{[32]} in refs. \cite{[20],[21]} and also
with the densities of the cluster-orbital shell-model
approximation (COSMA) \cite{[9],[10],[18],[19]}. In \cite{[21]} a
comparison of the obtained results has been performed to those
from the alpha-core approach with the complex and fully non-local
effective interaction \cite{[22]} and also with the non-core model
based on the large-scale shell-model (LSSM) calculations (refs.
\cite{[7],[28],[29]} and references therein). It was shown that
the elastic scattering is a good tool to distinguish between
different density distributions \cite{[21]}.  Usually, in the
usage of the complex OP's for analyses of the differential cross
sections, their imaginary part and the spin-orbit terms have been
determined in a phenomenological way, and then the OP's include a
number of fitting parameters. The question to optimize this number
in the analyses of the experimental data is not usually
considered.

The main aim of the present work is to calculate differential
cross sections of elastic $^6$He+p scattering at different
energies studying the possibility to describe the existing
experimental data by using a minimal number of fitting parameters.
We note that for this purpose we use the so-called high-energy
approximation (HEA) OP \cite{[33]}. Its form can be considered as
a microscopic folding of the densities of the colliding nuclei NN
scattering amplitude. There are no free fitting parameters in this
OP and its dependence on the energy is included in the input data
of the NN scattering amplitude and the total cross section. It is
generally believed that the Glauber (eikonal) approximation is
reasonable at energies of hundreds MeV and higher. However,
beginning from the work \cite{Vries80} the method of HEA has been
usually modified by replacing the eikonal straight-line trajectory
at an impact parameter $b$ by the parameter $b_{c}$ that
corresponds to the distance of the closest approach of the
projectile in the Coulomb plus the nuclear potential. For the
medium and heavy nuclei the Coulomb distortion dominates and
successful applications of the approach were demonstrated,
firstly, in \cite{Vitt87} and later on in many papers, $e.g.$ in
\cite{[33],Charagi90,Charagi97,Neto} at low ($>10$ MeV/N) and
intermediate energies to the description of the data on the
differential elastic and total reaction cross sections of various
projectile ions and target nuclei. The distortion effect caused by
the nuclear potential is important mainly for lightest nuclei as
shown in \cite{Brink81}. In the last decade such a modified
Glauber method turned out to be rather effective and was employed
in many works devoted to analyses of nucleus-nucleus scattering
processes. In the present paper, to avoid limitations in the
modified HEA formulae, we account for distortion effects by
computing cross sections using the DWUCK4 code of numerical
solving of the Schr\"{o}dinger equation. At the same time we use
(similarly to ref.~\cite{[51]}) the microscopic HEA imaginary part
of the OP obtained in ref.~\cite{[33]} that yields the same
eikonal phase as that given in the optical limit of the Glauber
microscopic theory of multiple scattering of complex systems. So,
one of the main aims of our work is to establish the limits of the
applicability of the HEA OP for calculations of differential
elastic cross sections of the $^{6}$He+p scattering for different
regions of angles and incident energies. Together with the HEA
imaginary potential we tested the OP whose form coincides with
that of the real part of the standard folding OP
\cite{[26],[27],[30]}. The latter includes an exchange term and,
correspondingly, the non-linear effects in calculations of the
potential. In the calculations we pay attention to the role of the
importance of the contribution of various physical quantities and
features, such as microscopically obtained spin-orbit forces and
regularization of the NN forces used in folding calculations.
Also, we consider effects of renormalization of the real and
imaginary parts of microscopic optical potentials, the differences
or similarities of various models for the $^6$He structure which
are used in the description of the experimental data on the cross
section of the $^6$He+p elastic scattering.

The theoretical scheme for microscopical calculations of the real
part of the OP's and cross sections is given in sect. 2. This
sect. includes also some methodical calculations. Section 3 is
devoted to the OP within the HEA. The results of the calculations
and the discussion are presented in sect. 4. Section 5 includes
the conclusions from the work.

\section{Basic relationships for calculations of the real part of the
nucleon-nucleus optical potential} \label{sec:2}
\subsection{Direct part of the real OP (Re OP)}
\label{sec:2.1} The real part of the nucleon-nucleus OP is assumed
to be a result of a single folding of the effective NN potential
with the nuclear density, $i.e.$ this is a particular case of the
double-folding \cite{[26]} in which a $\delta({\bf r}_1)$ function
has to be used for the density of the incoming particle $\rho({\bf
r}_1)$. Then the direct part of the Re OP ($V^D$) has the
following form of the isoscalar (IS)- and isovector (IV)-
contributions, correspondingly:
\begin{equation}\label{2.1}
V^D_{IS}(r)=\int \rho_2({\bf r}_2)g(E)F(\rho_2)v_{00}^D(s)d^3r_2,
\end{equation}
\begin{equation}\label{2.2}
V^D_{IV}(r)=\int \delta\rho_2({\bf
r}_2)g(E)F(\rho_2)v_{01}^D(s)d^3r_2,
\end{equation}
where ${\bf s}={\bf r}+{\bf r}_2$,
\begin{equation}\label{2.3}
\rho_2({\bf r}_2)=\rho_{2,p}({\bf r}_{2,p})+\rho_{2,n}({\bf
r}_{2,n}),
\end{equation}
\begin{equation}\label{2.4}
\delta\rho_2({\bf r}_2)=\rho_{2,p}({\bf r}_{2,p})-\rho_{2,n}({\bf
r}_{2,n}).
\end{equation}
Here $\rho_{2,p}({\bf r}_{2,p})$ and $\rho_{2,n}({\bf r}_{2,n})$
are the proton and neutron densities in the target nucleus. In
eqs. (\ref{2.1}) and (\ref{2.2}) $g(E)=1-0.003E$ represents the
energy dependence of the effective NN interaction while
$F(\rho_2)$ contains its density dependence. Following ref.
\cite{[30]} we use its form for the CDM3Y6 effective Paris
potential:
\begin{equation}\label{2.5}
F(\rho)=C\left [1+\alpha e^{-\beta\rho({\bf r})}-\gamma\rho({\bf
r})\right ],
\end{equation}
where $C=0.2658$, $\alpha=3.8033$, $\beta=1.4099$ fm$^3$,
$\gamma=4.0$ fm$^3$.

The effective NN interaction $v^D_{00(01)}$ in eqs. (\ref{2.1})
and (\ref{2.2}) has included the isoscalar and isovector
components of the direct part of the M3Y interaction based on the
results of the $g$-matrix calculations using the Paris NN
potential \cite{[27],[30]}. The M3Y potentials which are used in
the folding calculations of OP's are sums of Yukawa-type terms
$\exp(-\mu r)/ (\mu r)$. Using eqs. (\ref{2.1})-(\ref{2.5}) one
can obtain the following forms of the direct part of the isoscalar
Re OP expressed by integrals in the coordinate and momentum space,
correspondingly:
\begin{eqnarray}\label{2.6}
V^D_{IS}(r)&=&Cg(E)\int \Bigl[\rho_2({\bf r}_2) +\alpha
{\bar\varrho}_2({\bf r}_2)- \gamma{\tilde\varrho}_2({\bf
r}_2)\Bigr] \nonumber \\ & \times & v_{00}^D(s)d^3 r_2,
\end{eqnarray}
\begin{eqnarray}\label{2.7}
V^D_{IS}(r)&=&{Cg(E)\over 2\pi^2}\int_0^\infty
\Bigl[\rho_2(q)+\alpha {\bar\varrho}_2(q)-\gamma
{\tilde\varrho}_2(q)\Bigr] \nonumber \\ & \times &
v_{00}^D(q)j_0(qr)q^2 dq,
\end{eqnarray}
where ${\bar\varrho}_2({\bf r}_2), {\tilde\varrho}_2({\bf r}_2)$
and their Fourier transform have the forms:
\begin{equation}\label{2.8}
{\bar\varrho}_2({\bf r}_2)=\rho_2({\bf r}_2)
e^{-{\beta\rho}_2({\bf r}_2)},
\end{equation}
\begin{equation}\label{2.9}
{\tilde\varrho}_2({\bf r}_2)=\bigl[\rho_2({\bf r}_2)\bigr]^2,
\end{equation}
\begin{equation}\label{2.10}
\rho(q)=\int{e}^{i{\bf q}{\bf r}}\rho(r)d^3r= 4\pi\int_0^\infty
\rho(r)j_0(qr)r^2dr.
\end{equation}
Similarly, exchanging $\rho_2$ by $\delta\rho_2$ [eq. (\ref{2.4})]
one can obtain the isovector part $V^D_{IV}$ of the direct part Re
OP.

\subsection{Exchange part of the real OP (Re OP)}
\label{sec:2.2} In contrast to the case of the double-folding
potential where integration over the coordinates (${\bf r}_1$) of
the nucleons in the incoming nucleus takes place, in the case of
the nucleon-nucleus interaction one can obtain:
\begin{eqnarray}\label{2.11}
V^{EX}_{IS}(r)&=&g(E)\int \rho_2({\bf r}_2, {\bf r}_2-{\bf s})
F\left(\rho_2({\bf r}_2-{\bf s}/2)\right ) \nonumber \\ & \times &
v_{00}^{EX}(s) j_0(k(r)s)d^3 r_2.
\end{eqnarray}
For the density matrix $\rho_2({\bf r}_2, {\bf r}_2-{\bf s})$ in
eq. (\ref{2.11}) we use the approximation for the calculation of
the knock-on exchange term of the folded potential from
\cite{[34]} which preserves the first term of the expansion given
in \cite{[35]}:
\begin{equation}\label{2.12}
\rho_2\bigl({\bf r}_2, {\bf r}_2-{\bf s}\bigr)~\simeq~
\rho_2\bigl(|{\bf r}_2-{\bf s}/2|\bigr) {\hat
j}_1\bigl(k_{F,2}(|{\bf r}_2-{\bf s}/2|)\cdot s\bigr).
\end{equation}
In eqs. (\ref{2.11}) and (\ref{2.12}):
\begin{equation}\label{2.13}
{\hat j}_1(x)={3\over x}j_1(x)={3\over x^3}(\sin x-x\cos x),
\end{equation}
\begin{equation}\label{2.14}
F(\rho_2)=C\left[1+\alpha e^{-{\beta\rho_2({\bf r}_2-{\bf s}/2)}}-
\gamma\rho_2({\bf r}_2-{\bf s}/2)\right ].
\end{equation}
In our case  ${\bf r}+{\bf r}_2={\bf s}$, therefore  $|{\bf
r}_2-{\bf s}/2|=|{\bf r}-{\bf s}/2|=|{\bf x}_2|$ and making the
substitution d$^3{\bf r}_2$=d$^3{\bf s}$ at fixed ${\bf r}$, eq.
(\ref{2.11}) can be rewritten in the form:
\begin{eqnarray}\label{2.15}
V^{EX}_{IS}(r)&=&g(E)\int h_2({\bf r}-{\bf s}/2, {\bf s})
F\left(\rho_2(|{\bf r}-{\bf s}/2|)\right ) \nonumber \\
& \times & v_{00}^{EX}(s)j_0(k(r)s)d^3s,
\end{eqnarray}
where
\begin{equation}\label{2.16}
h_2({\bf x}_2,{\bf s})= \rho_2({\bf x}_2){\hat
j}_1\bigl(k_{F,2}({\bf x}_2)\cdot s\bigr), \qquad
\end{equation}
d$^3s=s^2$d$s \sin\theta $d$\theta $d$\phi $, with $\theta,~\phi$
being angles which determine the vector ${\bf s}/2$ with respect
to the fixed vector ${\bf r}$. Integrating over $d\phi$ gives the
factor $2\pi$. We separate the integral over d$x=\sin \theta
$d$\theta=-$d$\cos\theta$:
\begin{eqnarray}\label{2.17}
G_0^{IS}(r,s)&=&\int_{-1}^1 \rho_2\bigl(x_2(r,s,x)\bigr)\, {\hat
j}_1\bigl(k_{F,2}\left(x_2(r,s,x)\right )\bigr)\nonumber \\
& \times & F\bigl(\rho_2\left(x_2(r,s,x)\right)\bigr)\,dx,
\end{eqnarray}
where
\begin{equation}\label{2.18}
x_2(r,s,x)=|{\bf r}-{\bf s}/2|=\left[r^2+{s^2\over 4}+rsx\right
]^{1/2}.
\end{equation}
Finally, the isoscalar part of the exchange Re OP of the
nucleon-nucleus interaction has the form:
\begin{eqnarray}\label{2.19}
V^{EX}_{IS}(r)&=&2\pi g(E) \nonumber \\
& \times & \int G_0^{IS}(r,s) v_{00}^{EX}(s) j_0\left(k(r)s\right
) s^2ds.
\end{eqnarray}
In eqs. (\ref{2.11}) and (\ref{2.19}), $v^{EX}_{00}(s)$ is the
isoscalar part of the exchange contribution to the effective NN
interaction and the local momentum of the incident nucleon in the
field of the Coulomb and nuclear potential (Re OP) is \cite{[36]}:
\begin{equation}\label{2.20}
k^2=(2m/\hbar^2)\left[E_{c.m.}-Vc(r)-V(r)\right ]\left[{1+A_2\over
A_2}\right ].
\end{equation}
In eqs. (\ref{2.12}), (\ref{2.16}) and (\ref{2.17}) $k_{F,2}$
defines the average relative momentum \cite{[34],[36]}:
\begin{equation}\label{2.21}
k_{F,2}(r)=\left\{{5\over 3\rho}\left[\tau(\rho)- {1\over
4}{\bf\nabla}^2\rho(r)\right ]\right \}^{1/2},
\end{equation}
where we choose further the extended Thomas-Fermi approximation
\cite{[37],[38]} for the kinetic energy density
\begin{eqnarray}\label{2.22}
{\tau(\rho)\over 2}&\simeq&\tau_q(\rho_q)={3\over
5}\left(3\pi^2\right )^{2/3} \left[\rho_q(r)\right
]^{5/3}\nonumber \\
&+& {|{\bf\nabla}\rho_q(r)|^2\over 36\rho_q(r)}+
{{\bf\nabla}^2\rho_q(r)\over 3},
\end{eqnarray}
valid for each kind of particles $q=n,p$.

The isovector part of the exchange Re OP can be obtained from eqs.
(\ref{2.17}) and (\ref{2.19}) exchanging $\rho_2$ by
$\delta\rho_2$.

\subsection{``Regularization'' of the effective NN interaction in the form
of Yukawa potential} \label{sec:2.3} An important aspect of using
the M3Y effective NN potentials in their traditional form
\begin{equation}\label{2.23}
v(s)=\sum_j N_j{\exp(-\mu_j|s|)\over \mu_j |s|}
\end{equation}
is related to their regularization due to the singularities. The
latter aims to exclude the singularities at the point $|{\bf
s}|=0$, which have no physical meaning. Formally, this singularity
might affect the consideration of the nucleon-nucleon scattering,
but it does not cause any difficulties in the calculations of the
folding integral [$e.g.$ eqs. (\ref{2.6}) and (\ref{2.15})] to
generate the nucleon-nucleus and/or nucleus-nucleus potential.
Nevertheless, ``regularized'' M3Y potentials of the NN interaction
which do not contain the mentioned singularity have been used in
some works ($e.g.$ \cite{[21],[39]}) and in our work we would like
to consider this point in more details. Note that the physical
reason for that kind of problems is related to the breaking of the
meson theory at very short range due to the extended structure of
nucleons \cite{[40]}. That is why the one-boson-exchange
potentials are usually regularized by introducing, $e.g.$, the
monopole-, dipole-, exponential cut-off form factors \cite{[41]},
as well as that suggested in ref. \cite{[42]}. As shown in
\cite{[43]}, the most important feature of these form factors is
that they have the same principal range, and that the overall
results are insensitive to their detailed shape. In refs.
\cite{[21],[39]} cut-off form factor was taken to be equal to the
proton form factor $\rho_p(q)$, parametrized according to the
experimental data in \cite{[44]} as a sum of Gaussian functions.
As a test of the role of the regularization, in the present work
we take the same expression, and then use the corresponding
"smoothing" function in a coordinate space, which has a meaning of
the density distribution of the incoming proton
\begin{equation}\label{2.24}
\rho_p(r)=\sum_{i=1}^{3} a_i{1\over (\pi r_i^2)^{3/2}}\exp
\left(-{r^2\over{r_i^2}}\right ),
\end{equation}
where $\sum a_i=1$, $a_1$=0.506373, $a_2$=0.327922,
$a_3$=0.165705, $r_1^2=0.431566$ fm$^2$, $r_2^2=0.139140$ fm$^2$,
$r_3^2=1.525540$ fm$^2$, and $\langle r^2 \rangle$=0.77542 fm$^2$.
Then, the regularized NN potential is determined by
\begin{equation}\label{2.25}
v_{reg}(s_1)=\int \rho_p(r_1) v(s) d^3r_1, \qquad {\bf s}_1={\bf
r}_1+{\bf s}.
\end{equation}
Rewriting this integral in the momentum representation one gets
the expression used in \cite{[21],[39]} instead of $v(s)$
\begin{equation}\label{2.26}
v_{reg}(s_1)={1\over (2\pi)^3}\int \rho_p(q)\,v(q)\, {e}^{{i{\bf
q}{\bf s}_1}}d^3q,
\end{equation}
where the proton form factor
\begin{equation}\label{2.27}
\rho_p(q)=\int {e}^{{-i{\bf q}{\bf r}_1}}\rho_p(r_1) d^3r_1 =\sum
a_i \exp\left(-{q^2r_i^2\over4}\right )
\end{equation}
and
\begin{equation}\label{2.28}
v(q)=\sum_j N_j {4\pi\over \mu_j}{1\over \mu_j^2+q^2}.
\end{equation}
On the other hand, substituting the ``regularized'' $v_{reg}(s_1)$
potential (\ref{2.25}) in eq. (1) instead of $v(s)$ leads to the
expression
\begin{eqnarray}\label{2.29}
V^D_{IS}(r)&=&\int \rho_p({\bf r}_1) g(E) F(\rho) v^D_{00}(s)
\rho_2({\bf r}_2)d^3r_1d^3r_2,\nonumber \\
& &{\bf r}_1+{\bf s}={\bf r}+{\bf r}_2,
\end{eqnarray}
whose form coincides with the definition of the double-folding
potential of interaction of two colliding complex systems having
densities $\rho_p$ and $\rho_2$.

\subsection{Methodical calculations}
\label{sec:2.4} In this subsection we will present results of our
calculations of the $^6$He+p elastic scattering differential cross
sections studying the role of various factors, such as: i) the
choice of the density distribution of $^6$He, ii) effects of the
regularization, and iii) the effect of the spin-orbit term in its
dependence on the interaction potential.

In the calculations we use the following three density
distributions of $^6$He:

i)  the point-nucleon density
\begin{eqnarray}\label{2.30}
\rho^X_{point}&=&{2\over \pi^{3/2}}\biggl\{{1\over a^3}\exp
\left[-\left({r\over a}\right )^2\right ]\nonumber \\
&+&{1\over b^3}{X-2\over 3}\left({r\over b}\right )^2
\exp\left[-\left({r\over b}\right )^2\right ]\biggr\}
\end{eqnarray}
applied by Tanihata {\it et al.} \cite{[45]} for a comparison of
the measured total reaction cross section of $^6$He+$^{12}$C at
800 A  MeV with the respective expression from \cite{[46]}
obtained there in the optical limit of the Glauber theory. In
(\ref{2.30}) $X=Z,N$, and the parameter values of $a$ and $b$ are
determined from
\begin{equation}\label{2.31}
a^2={a^\ast}^2\left(1-{1\over A}\right ), \qquad
b^2={b^\ast}^2\left(1-{1\over A}\right ),
\end{equation}
where $a^*=1.53$ fm, $b^*=2.24$ fm, and hence $a=1.40$ fm, $b=
2.04$ fm. Thus, the rms radius of the point-proton density of
$^6$He is equal to 1.72 fm.

ii) the COSMA point-nucleon density \cite{[19]} which has the same
analytical form as eq. (\ref{2.30}), but with the parameter values
$a=1.55$ fm and $b= 2.24$ fm \cite{[9],[10]}, and the rms radius
of the point-proton density is equal to 1.89 fm.

We should emphasize that both Tanihata and COSMA densities have a
Gaussian asymptotic behavior which is not a realistic one at high
$q$. That is why we consider also other more realistic proton and
neutron densities of $^6$He:

iii) the LSSM densities obtained in a complete 4$\hbar\omega$
shell-model space \cite{[28]} using Woods-Saxon (WS)
single-particle wave function basis with realistic exponential\\
asymptotic behavior.

\begin{figure}[h]
\begin{center}
\resizebox{0.49\textwidth}{!}{%
\includegraphics{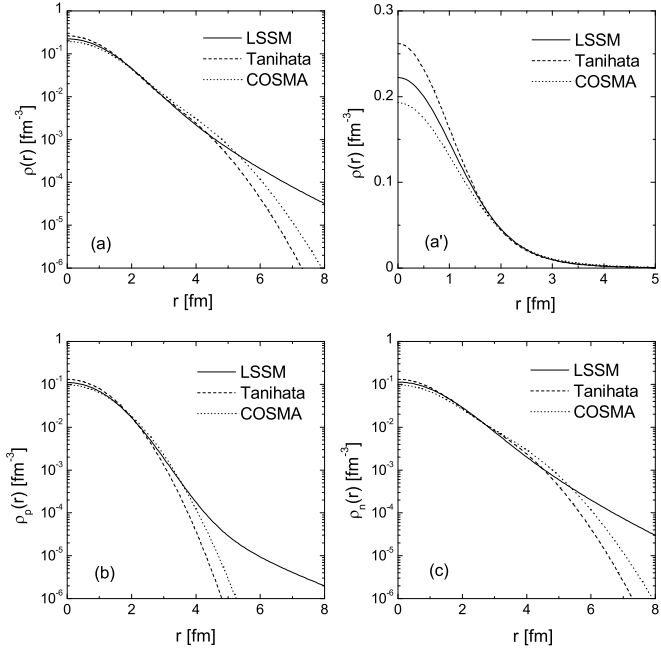}
} \caption{Total ((a) and (a$^\prime$)) , point-proton (b) and
point-neutron (c) densities from the model of Tanihata {\it et
al.} \cite{[45]}, COSMA \cite{[19]} and LSSM calculations
\cite{[28]}.} \label{fig:1}
\end{center}
\end{figure}

In fig. \ref{fig:1} we present the total density distribution of
$^6$He: $\rho(r)=\rho_p(r)+\rho_n(r)$ (in logarithmic and linear
scale), as well as the point-proton and point-neutron density
distributions of Tanihata {\it et al.}, COSMA and LSSM. One can
see that the Tanihata and COSMA densities have a Gaussian slope
while the LSSM tail occurs too higher and goes for the neutrons to
larger values of $r$ than for protons. As known, the differences
between the densities for smaller values of $r$ ($r<R$, $R$ being
the linear size of the nucleus) can be revealed mostly in the
nucleon-nucleus scattering, but in heavy ion collisions one can
study better the asymptotic region.

\begin{figure}[h]
\begin{center}
\resizebox{0.48\textwidth}{!}{%
\includegraphics{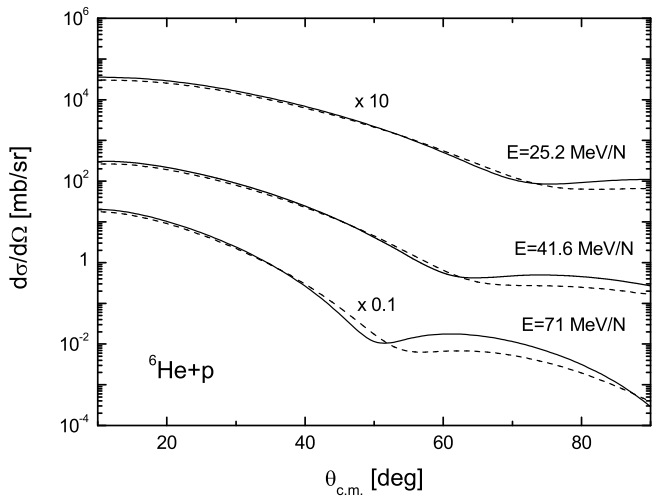}
} \caption{Elastic $^6$He+p scattering cross sections at different
energies computed by using the microscopically folded real OP and
the imaginary part taken in the same form ($V^F=V^D+V^{EX}$)
calculated using the LSSM density of $^6$He with (dashed line) and
without (solid line) regularization of the M3Y effective NN
interaction.} \label{fig:2}
\end{center}
\end{figure}

The role of the regularization (described in subsect. 2.3) is
shown in fig. \ref{fig:2} by a comparison of the $^6$He+p elastic
scattering differential cross sections at different energies:
25.2, 41.6, and 71 MeV/N. As can be seen, generally the effect of
the regularization is rather weak, but it increases when the
energy and the scattering angle increase. It starts to be seen at
smaller angles when the energy increases. Generally, checking the
role of the regularization we conclude that it is not necessary to
include it in the cases of folding OP's in the nucleon-nucleus
scattering.

\begin{figure}[h]
\begin{center}
\resizebox{0.46\textwidth}{!}{%
\includegraphics{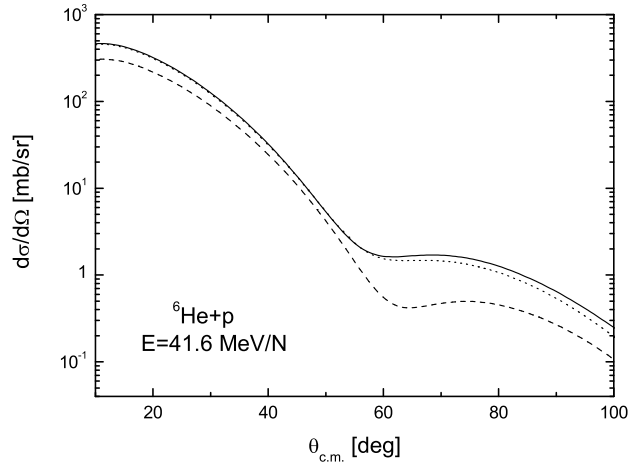}
} \caption{Elastic $^6$He+p scattering cross section at energy
$E=41.6$ MeV/N calculated using the microscopic Re OP ($V^F$) in
the spin-orbit term  $(1/r)$d$V^F/$d$r$ (solid line) and also
using the WS Re OP in $12.4 (1/r)$d$f^{WS}/$d$r$ (dotted line).
The dashed line shows calculations without spin-orbit term. The
LSSM density is taken for $^6$He and the Im OP has the form of
$V^F$.} \label{fig:3}
\end{center}
\end{figure}

In fig. \ref{fig:3} we show the role of the spin-orbit term in the
cross section of elastic $^6$He+p scattering at a given energy
$E=41.6$ MeV/N. The spin-orbit term is taken to be in the form:
\begin{equation}\label{2.32}
U_{so}\simeq N_{so}\lambda_\pi^2\left({1\over r}\right
){df(r)\over dr},
\end{equation}
where $f(r)$ is the form of the Re OP, and $\lambda_\pi^2$=2
fm$^2$. The calculations are given for three cases: i) when
$N_{so}=0.5$ and $f(r)$ is exchanged by the microscopic Re
OP=$V^{micro}(r)$ in MeV; ii) $N_{so}=6.2$ MeV and $f(r)$ is the
WS form factor taken from \cite{BG}, and iii) without spin-orbit
term, $i.e.$ when $N_{so}=0$.

One can see that the cross sections in the cases i) and ii) are
very close to each other. Our analysis showed that, generally,
there is not a strong dependence on the shape of $f(r)$. From the
other side, however, the effect of the spin-orbit term in the
considered case is important at larger angles ($>60^\circ$).

\section{Optical potential within the high-energy approximation}
\label{sec:3} As known, the real part of the optical potential can
be calculated microscopically in the standard form of the single-
or double-folded integral. In principle, the physical nature of
the real and imaginary parts of OP is different. The imaginary
part is related to the flux loss at the transition of the
particles from the elastic to the inelastic and reaction channels
which depends on both the structure of the colliding nuclei and
reaction mechanisms. This makes it difficult to construct
practically a convenient theory of the Im OP especially for
nucleus-nucleus scattering (see, $e.g.$, \cite{SW} and refs.
therein). However, in the case of more simple proton-nucleus
scattering one can get fairly good applications (see, $e.g.$,
\cite{GKB}) when using a single-folding pseudo-potential
\cite{[26]} multiplied by the fitted complex renormalization
factor $(N_R+iN_I)$ to obtain the complex potential with the unit
shape of the real and imaginary parts.  On the other hand, in the
case of heavy ion scattering, many applications were made when the
real part of OP was microscopically calculated while an imaginary
part was taken in the WS form with three or more fitted
parameters.

In this paper we intend to test the so-called HEA optical
potential (at least its imaginary part) to explain the available
data on the $^6$He+p differential cross sections at a few tens of
MeV/N. As already mentioned in the Introduction, the
generalization of the eikonal method made it possible to use it at
relatively low energies
\cite{[33],Vries80,Vitt87,Charagi90,Charagi97,Neto,Brink81}. In
our work, on the base of the eikonal phases we reconstruct the
equivalent potential including its imaginary part. The
calculations of the cross sections are performed {\it by numerical
integration of the Schr\"{o}dinger equation} by means of the code
DWUCK4 using all interactions obtained (Coulomb plus nuclear OP),
but not using the HEA scheme for cross section calculations. We
note that using this method we aim to establish {\it the limits of
the applicability of the imaginary part of the HEA optical
potential}. We also note that, firstly, in the calculations of the
HEA potential, similarly to ref.~\cite{[33],[51]}, we do not
neglect the parallel contribution of the transferred momentum (in
contrast to the Glauber method). As a result, the integral in the
expression for the potential includes spherical Bessel functions,
instead of the cylindrical ones (which appear in the expressions
for the eikonal HEA). Secondly, this makes softer the limitation
of the applicability of the method. We mention that in this case
the region of the applicability of the standard HEA for $E\gg
|U(\bar R)|$ and small angles is transformed to $E< |U(\bar R)|$
and $\vartheta<[\sqrt{2/k{\bar R}}+|U(\bar R)|/E]$, with $\bar R$
being the radius of the potential at its periphery ($e.g.$ at the
half-depth of the nuclear potential), where the absorption is
quite strong \cite{Luky2001}. Previously, the microscopic HEA OP
was successfully applied in refs. \cite{[33],ZLL,ZAL} to describe
the $^{16,17}$O+A elastic scattering data at about hundred of
MeV/N from \cite{Roussel,Neto}. Here we would like to give an
estimation for the angles of the application of our approach. For
the $^{6}$He+p scattering at $E_{lab}$=40 MeV/N and $\bar R\approx
$ 2 fm, it follows from the expression given above that
$\vartheta<50^\circ$, $i.e.$ this is the applicability region of
angles considered in our calculations. The basic component of HEA
is the eikonal phase $\Phi(b)$ which depends on the impact
parameter of the collision $b$. The amplitude and the cross
sections of scattering and reactions are expressed by means of
$\Phi(b)$. In the phenomenological approach the phase is given by
the integral along the axis $z$ of the interaction potential. On
the other side, this phase was derived in the microscopic level
for the proton-nucleus scattering in the Glauber theory
\cite{[47],[48]} and generalized later to the nucleus-nucleus
scattering in \cite{[49],[50]}. In the so-called ``optical limit''
of this theory an explicit form of the microscopical phase
$\Phi(b)$ is expressed by the densities of colliding nuclei and of
the amplitude of the NN scattering. By a comparison of both the
eikonal and microscopic expressions for the phase it became
possible in \cite{[33]} and \cite{[51]} to obtain the explicit
form of the OP in HEA, which gives a description of the
nucleus-nucleus scattering, being equivalent to that from the
microscopical approach:
\begin{equation}\label{3.1}
U^H_{opt}=V^H+iW^H,
\end{equation}
where
\begin{eqnarray}\label{3.2}
V^H(r)&=&-{\hbar v\over
(2\pi)^2}{\bar\sigma}_{NN}{\bar\alpha}_{NN}\nonumber \\
& \times & \int_0^\infty \rho_1(q) \rho_2(q) f_N(q) j_0(qr) q^2dq,
\end{eqnarray}
\begin{eqnarray}\label{3.3}
W^H(r)&=&-{\hbar v\over(2\pi)^2}{\bar\sigma}_{NN}\nonumber \\
& \times & \int_0^\infty \rho_1(q) \rho_2(q) f_N(q) j_0(qr) q^2dq,
\end{eqnarray}
with $v$ being the velocity of the nucleus-nucleus relative
motion, $\rho_{1,(2)}(q)$ being the form factors corresponding to
the point-like nucleon density distributions of the nuclei and
$f_N(q)$ being the amplitude of the NN scattering which depends on
the transfer momentum $q$ (see, $e.g.$ \cite{Alkhaz78}). The
quantities ${\bar\sigma}_{NN}$ and ${\bar\alpha}_{NN}$ are the
averaged over the isospins total NN cross section and the ratio of
the real to imaginary part of the scattering amplitude at zero
angle of free nucleons. They can be obtained from the data on the
mutual scattering of nucleons. For instance, the parametrizations
of the energy dependence of ${\bar\sigma}_{NN}$ \cite{Charagi90}
and ${\bar\alpha}_{NN}$ in the interval $\epsilon_{lab}$ = 10
MeV$\div$ 1 GeV \cite{[53]} are known:
\begin{equation}\label{3.4}
{\bar\sigma}_{NN}=\frac{Z_1Z_2 \sigma_{pp}+N_1N_2 \sigma_{nn}+
\zeta \sigma_{np}}{A_1A_2},
\end{equation}
$\zeta=Z_1N_2+N_1Z_2$,
\begin{eqnarray}\label{3.5}
\sigma_{np}&=&\left(-70.67-18.18 \beta^{-1}+25.26
\beta^{-2}+113.85 \beta\right )\nonumber \\
& \times &  f_m(np),
\end{eqnarray}
\begin{eqnarray}\label{3.6}
\sigma_{pp}=\sigma_{nn}&=& \left(13.73 - 15.04 \beta^{-1} + 8.76
\beta^{-2} + 68.67 \beta^4\right )\nonumber \\
&\times &f_m(nn),
\end{eqnarray}
where
\begin{equation}\label{3.7}
\beta=\frac{v}{c}=\sqrt{1-\Bigl({931.5\over
\varepsilon_{lab}+931.5}\Bigr)^2}
\end{equation}
is the ratio of the relative to the light velocities,
$\epsilon_{lab}=E/A_1$ is the energy (in MeV) of a nucleon in the
incident nucleus in the laboratory system \footnote{In our case
one has $Z_1=A_1=1, N_1=0$.}, and cross sections are given in mb.
The factors $f_m(np)$ and $f_m(nn)$ are introduced to correct the
dependence of cross sections on the energy and on the density of
nuclear matter. The in-medium NN interaction has been widely
investigated. For instance, in \cite{[54]} numerical calculations
of the total NN cross sections have been performed on the basis of
the Dirac-Brueckner theory of nuclear matter and their
parametrization has been given in ref. \cite{[55]} to obtain the
following correcting factors:
\begin{equation}\label{3.8}
f_m(np)=\frac{1+20.88 \varepsilon_{lab}^{0.04} \rho^{2.02}}
{1+35.86 \rho^{1.90}},
\end{equation}
\\
\begin{equation}\label{3.9}
f_m(nn)=\frac{1+7.772 \varepsilon_{lab}^{0.06} \rho^{1.48}}
{1+18.01 \rho^{1.46}}.
\end{equation}
In eqs. (\ref{3.8}) and (\ref{3.9}) the densities are in
fm$^{-3}$. In the case of free nucleons ($\rho$=0)
$f_m(np)$=$f_m(nn)$=1. The increase of the density leads to
decrease of these factors and, correspondingly, of the cross
sections.

Here we present also the ratio of the real to imaginary part of
the NN amplitude at zero angle averaged over the nuclear isospins
and parametrized in \cite{[53]}:
\begin{equation}\label{3.10}
\bar\alpha_{NN}=\frac{Z_1Z_2\alpha_{pp}\sigma_{pp}+
N_1N_2\alpha_{nn}\sigma_{nn}+\zeta\alpha_{np}\sigma_{np}}
{Z_1Z_2\sigma_{pp}+N_1N_2\sigma_{nn}+\zeta\sigma_{np}},
\end{equation}
\begin{equation}\label{3.11}
\alpha_{nn}=\alpha_{pp}=0.0078+0.1762\sqrt{\varepsilon_{lab}}+
0.01436\varepsilon_{lab},
\end{equation}
\begin{equation}\label{3.12}
\alpha_{np}=-0.0301+0.2148\sqrt{\varepsilon_{lab}}-0.0551\varepsilon_{lab}.
\end{equation}

\begin{figure}[ht]
\begin{center}
\resizebox{0.45\textwidth}{!}{%
\includegraphics{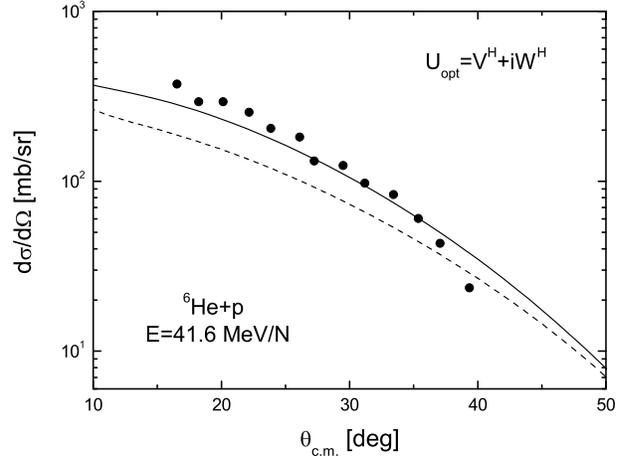}
} \caption{Elastic $^6$He+p scattering cross section at energy
$E=41.6$ MeV/N calculated by using $U_{opt}=V^H+iW^H$ with (solid
line) and without (dashed line) spin-orbit term
$(1/r)$d$V^H/$d$r$. The $^6$He LSSM density is used. Experimental
data are from refs. \cite{[6],[8]}.} \label{fig:4}
\end{center}
\end{figure}

The results of calculations of the $^6$He+p elastic scattering
cross sections at energy 41.6 MeV/N using the OP in the form
(\ref{3.1}) are presented in fig. \ref{fig:4} with and without
spin-orbita term of the form of
0.5$\lambda_\pi^2($d$/$d$r)V^{micro}$. It is seen a better
agreement with the data up to $\vartheta\simeq 40^\circ$ in the
case when the spin-orbit term is included. The LSSM densities of
$^6$He have been used in the calculations. Here we would like to
mention the question whether the data can be reasonably fitted
either by introducing a spin-orbit force, or by renormalizing the
potential. For the folding potential these two ways are not
equivalent because of the role of the spin-orbit term at large
angles. However, in the case of the HEA, responsible for the area
of comparably small angles, it seems likely that one could fit the
data using either a spin-orbit term or a renormalization. We note,
however, that our main goal is to show that the HEA OP's can be
used in their domain of validity with no additional free
parameters; this is certainly true with the inclusion of the
spin-orbit term at angles smaller than 40$^\circ$.

In the next sect., following the basic theoretical scheme given in
sects. 2 and 3, we present the results of our calculations. For
calculations of Re OP we use the effective M3Y interaction based
on the Paris NN potential with a density dependence in the form
CDM3Y6 from \cite{[30]} without ``regularization''. Also we
account for the spin-orbit term $\sim $d$V^{micro}/$d$r$ and use
the LSSM proton and neutron densities of $^6$He. Besides, we
include in the calculations the imaginary part of the OP's
obtained within the HEA.

\section{Results of calculations and discussion}
\label{sec:4} In this sect. we present our calculations of the
cross sections of $^6$He+p elastic scattering at different
energies aiming to study: i) effects of the different behavior of
the model densities of $^6$He; ii) the possibility to use the HEA
optical potentials for different energies and angles and also the
microscopic OP with the folded real part $V^F=V^D+V^{EX}$ and the
HEA imaginary part $W^H$; iii) the role of the renormalization of
the depths of the real and imaginary parts of OP's, and iv) the
in-medium effects of the effective NN interaction on the
microscopically calculated Re OP.

We start this sect. with calculations of the $^6$He+p elastic
cross sections using the real part of OP calculated within the
folding approach ($V^F$), as well as the real ($V^H$) and
imaginary ($W^H$) parts of the HEA OP's. We introduce two
renormalization parameters $N_R$ and $N_I$ (already mentioned
above) and consider the following three types for the OP's:
\begin{equation}\label{4.1}
(A) \hspace*{1cm}   U^A_{opt}\,=\,N^A_{R}V^H\,+\,iN^A_{I}W^H,
\end{equation}
\begin{equation}\label{4.2}
(B) \hspace*{1cm}   U^B_{opt}\,=\,N^B_{R}V^F\,+\,iN^B_{I}W^H,
\end{equation}
\begin{equation}\label{4.3}
(C) \hspace*{1cm}   U^C_{opt}\,=\,N^C_{R}V^F\,+\,iN^C_{I}V^F.
\end{equation}

As can be seen, in case (A) we use both real and imaginary parts
from the HEA calculations  of OP's; in case (B) we take $V^F$, the
folded real part of the microscopic OP where the exchange term is
included, and the imaginary part is applied in the form
(\ref{3.3}) of the HEA OP, while in case (C) we use the
microscopically folded real and imaginary parts in the form of
$V^F$, and thus they have the same shape.

\begin{figure}[h]
\begin{center}
\resizebox{0.38\textwidth}{!}{%
\includegraphics{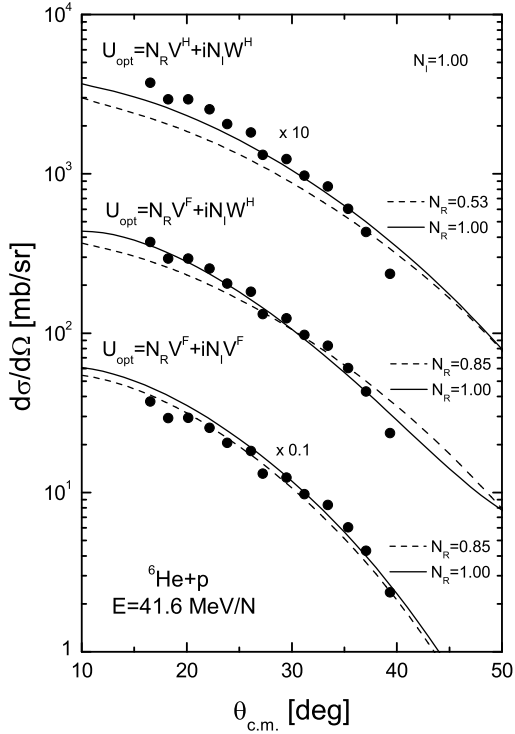}
} \caption{Elastic $^6$He+p scattering cross section at energy
$E=41.6$ MeV/N calculated using different OP's (cases (A), (B) and
(C) from the text) for various values of the renormalization
parameters $N_R$ ($N_I=1$). The LSSM density of $^6$He is applied.
Experimental data are from refs. \cite{[6],[8]}.}
\label{fig:5}
\end{center}
\end{figure}

In fig. \ref{fig:5} are presented the results of calculations of
the $^6$He+p elastic cross sections (for energy $E$=41.6 MeV/N)
with the fixed value N$_{I}$=1 for all three cases (A), (B) and
(C). The comparison with the data is performed for two values of
$N_R$ for each case, namely 0.53 and 1.00 for case (A), 0.85 and
1.00 for cases (B) and (C). One can see a good agreement with the
data using $N_R$=1.00 for the cases (A) and (B) and using $N_R$ =
0.85 for case (C). It can be concluded that for both cases (A) and
(B) the renormalization is not necessary. As in the case shown in
fig. \ref{fig:4}, we note that the agreement obtained by using the
HEA is for angles smaller than 40$^\circ$.

\begin{figure}[h]
\begin{center}
\resizebox{0.46\textwidth}{!}{%
\includegraphics{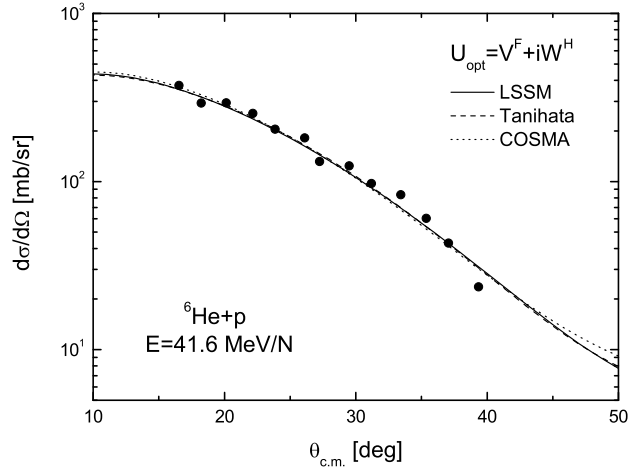}
} \caption{Elastic $^6$He+p scattering cross section at energy
$E=41.6$ MeV/N calculated using $U_{opt}=V^F+iW^H$ and Tanihata
(dashed line), COSMA (dotted line) and LSSM (solid line) densities
of $^6$He. Experimental data are from \cite{[6],[8]}.}
\label{fig:6}
\end{center}
\end{figure}

The latter concerns also the results of the calculations of
$^{6}$He+p elastic scattering ($E$=41.6 MeV/N) in case (B) using
three densities of $^6$He given by Tanihata {\it et al.}, COSMA
and from the LSSM given in fig. \ref{fig:6}. It can be seen a good
agreement with the empirical data up to $\vartheta\simeq
35^\circ$. Deviations of the results with the use of the COSMA
density from the other two cases start at angles larger than
40$^\circ$. In these calculations $N_R=N_I=1$.

\begin{figure}[h]
\begin{center}
\resizebox{0.5\textwidth}{!}{%
\includegraphics{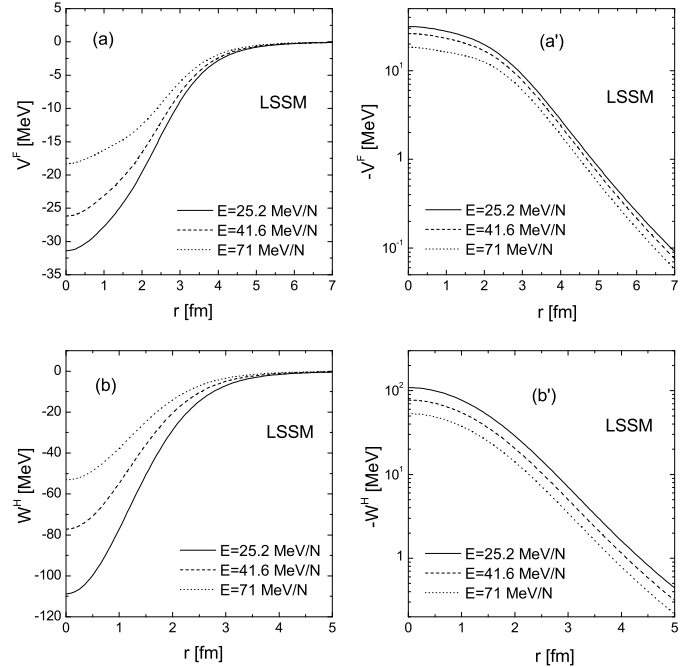}
} \caption{Microscopic real part ($V^F$) of OP ((a) and
(a$^\prime$)) and HEA imaginary part ($W^H$) ((b) and
(b$^\prime$)) calculated using the LSSM density of $^6$He for
energies $E=25.2$ (solid lines), 41.6 (dashed lines) and 71 MeV/N
(dotted lines).} \label{fig:7}
\end{center}
\end{figure}

In fig. \ref{fig:7} we present the real (V$^F$) and imaginary
(W$^H$) parts of the OP's (case (B)) calculated  using the LSSM
density of $^6$He for three different energies: 25.2, 41.6, and 71
MeV/N. They are given without renormalization ($i.e.$
$N_R$=$N_I$=1) and are illustrated in logarithmic and linear
scales. One can see the decrease of the potential depths with the
increase of the energy. Here we would like to note that, as can be
seen from fig. \ref{fig:7}, at ${\bar R}\approx 2$ fm the values
of the potentials for $E_{lab}=40$ MeV/N are $U({\bar R})\approx
15-20$ MeV, and thus the limit condition for the HEA $E>U({\bar
R})$ is fulfilled. This explains the applicability of the HEA at
such values of the energy. We should mention the larger depth of
the HEA imaginary parts $(W^H)$ of the OP for the case of 25.2
MeV/N seen in fig. \ref{fig:7}. As we will see below, the result
of the HEA for $W^H$ in this case affects significantly the cross
section.

\begin{figure}[h]
\begin{center}
\resizebox{0.31\textwidth}{!}{%
\includegraphics{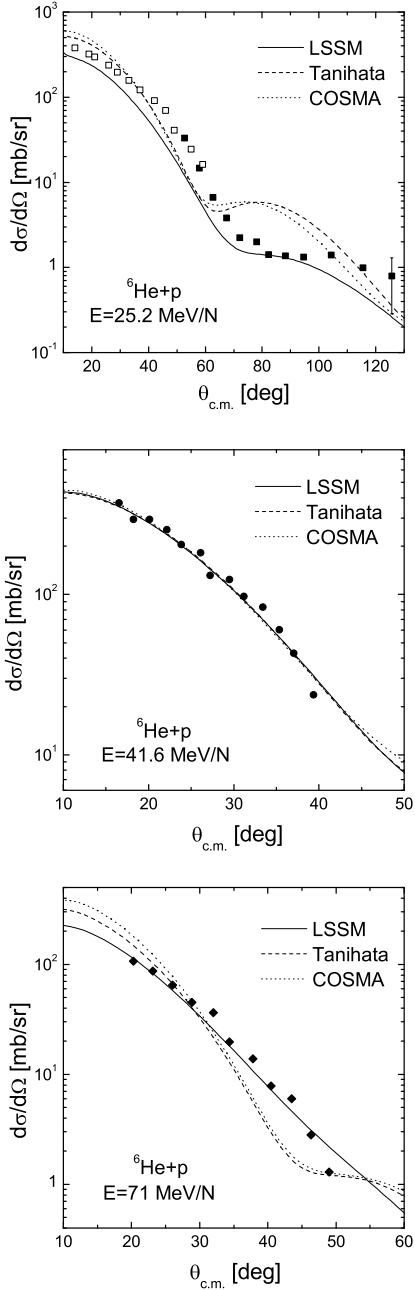}
} \caption{Elastic $^6$He+p scattering cross sections at different
energies calculated using $U_{opt}=N_{R}V^{F}+iN_{I}W^{H}$ for
various values of the renormalization parameters $N_R$ and $N_I$
giving a reasonable agreement with the data (presented in table
\ref{tab:1}). The used densities of $^6$He are LSSM (solid line),
Tanihata (dashed line) and COSMA (dotted line). Experimental data
are taken for 25.2 \cite{[1],[2],[3]}, 41.6 \cite{[6],[8]} and 71
MeV/N \cite{[9],[10]}.} \label{fig:8}
\end{center}
\end{figure}

The three densities are used to calculate the cross sections of
$^6$He+p elastic cross sections for three energies (25.2, 41.6,
and 71 MeV/N) shown in fig. \ref{fig:8}. One can see the fairly
good agreement with the experimental data of the results with the
LSSM density for $^6$He for energies 41.6 and 71 MeV/N, in
contrast to the results obtained with the other two densities for
the energies 25.2 and 71 MeV/N.
In table \ref{tab:1} we list the values of the renormalization
parameters $N_R$ and $N_I$ that give a reasonable agreement with
the data for the three energies and the three different densities
of $^6$He shown in fig. \ref{fig:8}. One can see also the
corresponding depths $N_R V_F^{(r=0)}$ and $N_I W_H^{(r=0)}$ of
the real and imaginary parts of OP's. The values of $N_R$ and
$N_I$ were chosen starting from the values $N_R=1$ and $N_I=1$ and
decreasing them gradually in order to achieve a reasonable fit to
the experimental data. In our opinion, the obtained values of
$N_R$ and $N_I$ still do not reveal some regular change with the
increase of the energy.

\begin{table*}
\caption{The optimal values of the renormalization parameters
$N_R$ and $N_I$ obtained by fitting the experimental data for the
elastic $^6$He+p cross sections. In the calculations $U_{opt}=N_R
V^F+i N_I W^H$ and LSSM, Tanihata and COSMA densities for energies
$E$=25.2, 41.6, and 71 MeV/N are used (the results are shown in
fig. \ref{fig:8}). The depths of the corresponding potentials (in
MeV) are presented, as well.}
\label{tab:1}       
\begin{center}
\begin{tabular}{lllllll}
\hline\noalign{\smallskip}
Energy   & 25.2 & 25.2 & 41.6 & 41.6 & 71 & 71 \\
\noalign{\smallskip}\hline\noalign{\smallskip}
Density  & $N_R$ & $N_I$ & $N_R$ & $N_I$ & $N_R$ & $N_I$ \\
\noalign{\smallskip}\hline\noalign{\smallskip}
LSSM     &  0.6  &  0.8  &  1.0  &  1.0  &  0.6  &  1.0  \\
Tanihata &  1.0  &  0.6  &  1.0  &  1.0  &  1.0  &  0.5  \\
COSMA    &  1.0  &  0.6  &  1.0  &  1.0  &  0.8  &  1.0  \\
\noalign{\smallskip}\hline\noalign{\smallskip}
Energy   & 25.2 & 25.2 & 41.6 & 41.6 & 71 & 71 \\
\noalign{\smallskip}\hline\noalign{\smallskip} Density  &
$N_RV_F^{(r=0)}$ & $N_IW_H^{(r=0)}$ & $N_RV_F^{(r=0)}$ &
$N_IW_H^{(r=0)}$ & $N_RV_F^{(r=0)}$ & $N_IW_H^{(r=0)}$ \\
\noalign{\smallskip}\hline\noalign{\smallskip}
LSSM     & 18.86 & 87.02 & 26.22 & 77.20 & 11.01 & 53.09 \\
Tanihata & 30.82 & 39.15 & 25.54 & 46.31 & 17.56 & 15.92 \\
COSMA    & 29.70 & 30.05 & 24.73 & 35.54 & 13.78 & 24.44 \\
\noalign{\smallskip}\hline
\end{tabular}
\end{center}
\end{table*}

\begin{figure}[h]
\begin{center}
\resizebox{0.45\textwidth}{!}{%
\includegraphics{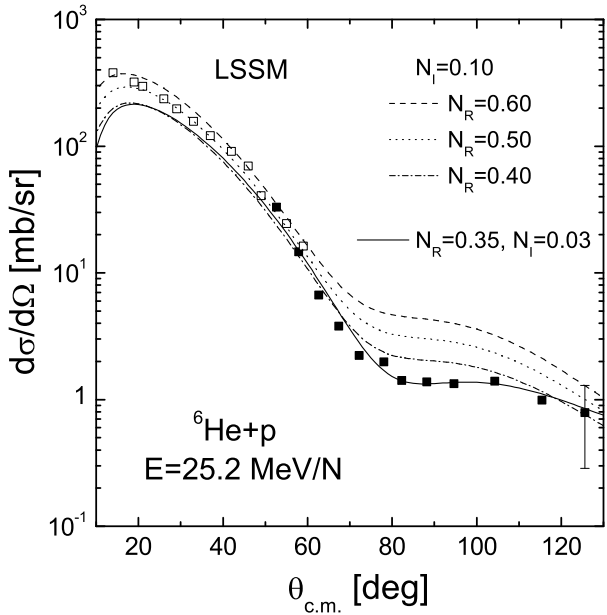}
} \caption{Elastic  $^6$He+p scattering cross sections for
$E=25.2$ MeV/N calculated by using the LSSM density for $^6$He.
The curves exhibit results for $U_{opt}=N_R V^F+i N_I W^H$ with
different values of $N_R$ (0.6-dashed, 0.5-dotted,
0.4-dash-dotted) and fixed value of $N_I$=0.1. The solid curve is
for $N_R$=0.35 and $N_I$=0.03. The experimental data
\cite{[1],[2],[3]} are also given.} \label{fig:9}
\end{center}
\end{figure}

In fig. \ref{fig:9} a particular attention is paid to the case of
the elastic $^6$He+p cross section at energy of 25.2 MeV/N
calculated using the LSSM density for the $^6$He nucleus. One can
see that, in contrast to the case of larger energies, in this case
smaller values of $N_R$ (0.35) and especially of $N_I$ (0.03) are
necessary for a better agreement with the experimental data. This
concerns the slope of the  cross section for angles
$\vartheta_{c.m.}$ between 70$^\circ$ and 120$^\circ$. We note
that the results shown in fig. \ref{fig:9} for the energy 25.2
MeV/N are for angles up to $\vartheta_{c.m.}\approx 120^\circ$.
The necessity to use much smaller value of the renormalization
parameter $N_I$ $(N_I=0.03)$ to fit the data for large angles is
related to the large value of the depth of the imaginary part of
the OP obtained within the HEA $(W^H)$ for this case, as already
mentioned above. This already shows the limitation of the approach
(the case (B), Eq.~(\ref{4.2})) for small energies ($<25$ MeV/N)
and large angle values. In this case the condition $E>|U({\bar
R})|$ is not already fulfilled. Along this line we should mention
the Ref.~\cite{Gupta2005} in which a microscopical  pseudo-folding
potential (without accounting for the exchange)
$V_{micro}=(N_R+iN_I)V_{folding}$ was used for calculations of
p+$^{18}$Ne and p+$^{18}$O scattering at energies 24.5 and 30
MeV/N. It was shown a good agreement with the data with $N_I=0$
for $^{18}$Ne and $N_I=0.006$ for $^{18}$O. These results confirm
the necessity to use microscopical rather than phenomenological
OP's also for low energy scattering.

\begin{figure}[h]
\begin{center}
\resizebox{0.46\textwidth}{!}{%
\includegraphics{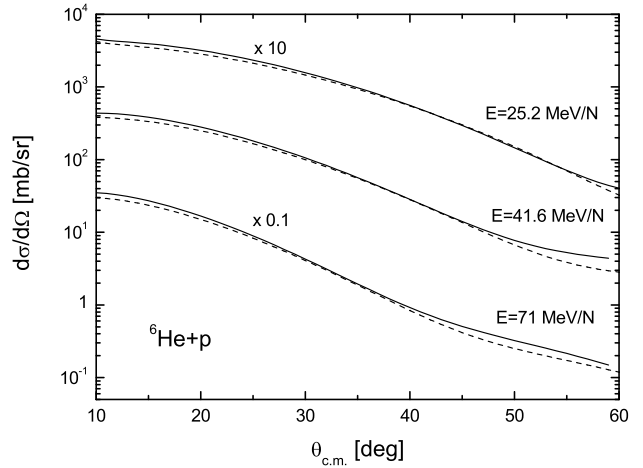}
} \caption{In-medium effect of the M3Y NN interaction on
calculations of elastic $^6$He+p scattering cross sections for
different energies. The Re OP ($V^F$) is calculated with
$F(\rho)=C(1+\alpha\exp(-\beta\rho)- \gamma\rho)$ (solid lines)
and with $F(\rho)=1$ (dashed lines). The imaginary part is
calculated within the HEA ($W^H$).}
\label{fig:10}
\end{center}
\end{figure}

In fig. \ref{fig:10} we present the in-medium effect of the NN
interaction on the elastic scattering cross sections of $^6$He+p.
The calculations were performed using the function $F(\rho)$ in
the form (\ref{2.5}) and with $F(\rho)$=1 (the case without
in-medium effect). The effective M3Y interaction based on the
Paris NN potential and the LSSM density of $^6$He were used in the
calculations. One can see the existence of small differences
between the results using density dependent and independent
effective NN interaction. The differences for the larger angles
increase with the energy increase.

\begin{figure}[htb]
\begin{center}
\resizebox{0.44\textwidth}{!}{%
\includegraphics{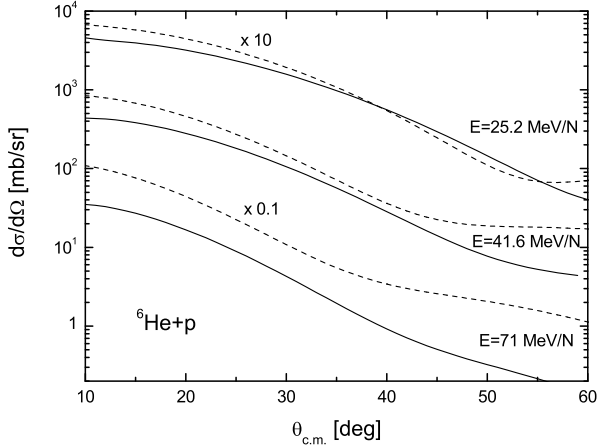}
} \caption{The non-linearity effect of elastic $^6$He+p scattering
cross section for different energies. The solid lines are the
calculation results when Re OP ($V^F$) includes $j_0(k(r)\cdot s)$
term in eq. (11) and the dashed lines are without this term. The
LSSM density of $^6$He is used. The imaginary part of OP is
calculated within the HEA ($W^H$).} \label{fig:11}
\end{center}
\end{figure}

Also, we studied the non-linear effects in calculations of the
exchange part of OP. As can be seen, the factor $j_0(k(r)\cdot s)$
takes place in the expressions for the $V^{EX}$ (see eqs.
(\ref{2.11}), (\ref{2.15}) and (\ref{2.19})), where the local
momentum of the relative motion $k(r)$ [eq. (\ref{2.20})] is
expressed by the OP. The calculations need an iteration procedure
and this makes the task complicated. However, if $k(r)\simeq 0$,
then $j_0\simeq 1$ and this simplifies calculations. In fig.
\ref{fig:11} we present the results for the $^6$He+p elastic
scattering cross section at different energies with and without
accounting for the factor $j_0(k(r)\cdot s)$ in the above
mentioned equations. As can be seen, the non-linearity effect is
small for the energy $E=25.2$ MeV/N, but it increases with the
energy increase up to an order of magnitude for $\vartheta >$
40$^\circ$ for the energy $E=71$ MeV/N. These results show the
necessity to perform calculations accounting for the non-linearity
of the task.

\section{Conclusions}
\label{sec:5} The results of the present work can be summarized as
follows:

i) The optical potentials and cross sections of $^6$He+p elastic
scattering were calculated at three different energies $E$ = 25.2,
41.6, and 71 MeV/N. The following components of the OP's were
used:

a) the real part of the OP (V$^F$) calculated microscopically
using the folding procedure and M3Y effective interaction based on
the Paris NN potential;

b) the real (V$^H$) and imaginary (W$^H$) parts of the OP
calculated within the high-energy approximation (HEA);

c) three different combinations of V$^F$, V$^H$ and W$^H$ (cases
(A), (B) and (C), eqs. (\ref{4.1})-(\ref{4.3})) were used for the
OP U$_{opt}$ in calculations of the elastic $^6$He+p cross
sections. The renormalization parameters $N_R$ and $N_I$ have been
introduced and their role has been studied.The cross sections were
calculated by numerical integration of the Schr\"{o}dinger
equation by means of DWUCK4 code using all interactions obtained
(Coulomb plus nuclear optical potential);

d) three different model densities of protons and neutrons in
$^6$He were used in the calculations: the phenomenological ones in
the form (\ref{2.30}), parametrized by Tanihata {\it et al.}, the
same form (\ref{2.30}) with parameters from the COSMA, and also
the microscopically calculated density within the LSSM.

ii) The results of our calculations show that the LSSM density of
$^6$He is the most preferable one because it leads to a better
agreement with the data for the $^6$He+p elastic scattering at the
three energies. The physical reason for the latter is that the
LSSM densities have more diffuse tails at larger $r$ than the
densities based on Gaussians. With the LSSM density and with the
optical potentials of the form (B), namely $U_{opt}= N_R V^F+iN_I
W^H$, we tried to choose the parameters $N_R$ and $N_I$ starting
from the values $N_R=1$ and $N_I=1$ and decreasing them gradually
in order to achieve a reasonable agreement of the calculated cross
sections with the available data. The obtained set of these
parameters is the following one: $N_R$ = 0.6, 1.0 and 0.6 and
$N_I$ = 0.8, 1.0 and 1.0 for energies 25.2, 41.6, and 71 MeV/N,
respectively. We note that the use of the microscopic folding real
part $V^F$ and of the HEA imaginary part $W^H$ leads to a good
agreement with the data for 41.6 and 71 MeV/N, while the data at
lowest energy 25.2 MeV/N are explained only on the qualitative
level. As shown by the estimations presented, this is related to
the limitations of using the imaginary part of the HEA OP for
energy smaller than around 25 MeV/N due to the fact that the
potentials do not fulfill the applicability condition $E>|U({\bar
R})|$. In this case the large value of the depth of the Im OP
obtained in the HEA $W^H$ has to be strongly reduced ($e.g.$ using
in our case $N_I=0.03$) in order to achieve a reasonable agreement
with the data for the energy 25.2 MeV/N.

iii) It was shown that the effect of the regularization of the M3Y
NN effective interaction is rather weak (with a small increase
with the increase of the energy and the angle) and, our conclusion
is that it is not necessary to use it applying the folding
approach to the cases of nucleon-nucleus scattering.

iv) The results show that the spin-orbit interaction is rather
important, particularly at angles larger than 60$^\circ$, and that
one can use in the $ls$-term the microscopically calculated Re OP
instead of the phenomenological WS potential with the three fitted
parameters.

v) The study of the dependence of the effective M3Y NN forces on
the nuclear matter density shows small differences between optical
potentials calculated with and without inclusion of the in-medium
effect. The difference between the corresponding cross sections
appears at larger angles and increases with the energy increase.

vi) We showed that the effect of the non-linearity on calculations
of Re OP connected with the factor $j_0(k(r)\cdot s)$, where
$k(r)$ is the local momentum of motion, is small for the energy of
25.2 MeV/N but it increases with the energy increase up to an
order of magnitude for $\vartheta_{c.m.}>40^\circ$ for the energy
$E=71$ MeV/N. Thus, the non-linearity in the calculations of the
Re OP should be taken into account in the calculations.

Concluding, we would like to note that it follows from our results
that the OP's can be calculated in the form $U^B_{opt} = N_R
V^F+iN_I W^H$ ($i.e.$ with microscopically calculated folding real
part ($V^F$) and with calculated within HEA imaginary part
($W^H$)) using only two free parameters ($N_R$ and $N_I$) which
renormalize the depths of the real and imaginary parts of OP.
Thus, it is not necessary (at least on the basis of the existing
experimental data on $^6$He+p elastic scattering cross sections)
to introduce a large number of fitting parameters, as is usually
done in the arbitrarily chosen forms of the phenomenological and
semi-microscopic optical potentials. It was pointed out that our
approach can be applied to cases with energies smaller than 100
MeV/N, like those of 71 MeV/N and 41.6 MeV/N considered in the
present work. It can be concluded that microscopical optical
potentials (including their imaginary part, $e.g.$ of the
HEA-type) rather than phenomenological ones have to be applied
also for the cases of tens of MeV/N. Thus the approach can be used
along with other more sophisticated methods like that from the
microscopic $g$-matrix description of the complex proton optical
potential and others.

\begin{acknowledgement}
The work is partly supported on the basis of the Project from the
Agreement for co-operation between the INRNE (Sofia) and JINR
(Dubna). Two of the authors (A.N.A. and M.K.G.) are grateful for
the support of the Bulgarian Science Fund under Contracts Nos.
$\Phi$--1416 and $\Phi$--1501. The author M.K.G. is grateful for
the warm hospitality given by the CSIC and for support during his
stay there from the State Secretariat of Education and
Universities of Spain (N/Ref. SAB2005--0012). The authors E.V.Z.
and K.V.L. thank the Russian Foundation for Basic Research (Grant
No. 06-01-00228) for the partial support.
\end{acknowledgement}

%

\end{document}